\documentclass{ringb99}
\usepackage{graphics}

\begin{document}
\title{Radio Observations of Merging Clusters in the Shapley Concentration}
\author{T. Venturi\inst{1}, S. Bardelli\inst{2}, G. Zambelli\inst{3},
R. Morganti\inst{1}
 \and R.W. Hunstead\inst{4}}  
\institute{Istituto di Radioastronomia, CNR, Via Gobetti 101, 40129 Bologna,
Italy \and Osservatorio Astronomico di Bologna, Via Ranzani 1, 40127 Bologna,
Italy \and Dipartimento di Astronomia, Universit\`a di Bologna,
Via Ranzani 1, 40127 Bologna, Italy
\and School of Physics, University of Sydney, NSW 2006, Australia}
\authorrunning{Venturi et al.}
\titlerunning{Radio Observations of Merging Clusters}
\maketitle

\begin{abstract}

In this paper we present the first results of a radio
survey at 22 cm in the central region of the Shapley Concentration,
carried out with the Australia Telescope Compact
Array.
In order to study the effect of merging on
the statistical properties of radio galaxies, and the relation
between merging and the existence of {\it relic} and {\it halo} type
radio sources, we observed the two complexes of merging clusters centered
on A3528 and on A3558.

Our results show that the radio source counts in these regions,
characterised by very high optical overdensity and 
major cluster mergers, do not differ from
the background source counts. This suggests that the merging phenomenon
does not influence the probability of a galaxy to become radio
surce.

Furthermore we investigated the possibility that the
extended radio source J1324$-$3138 is a relic, applying the model
recently proposed by Ensslin et al. (1998). Our
analysis shows that the properties of J1324$-$3138 are consistent
with the idea that the source is a relic located on the shock front
between A3556 and a smaller group accreting onto the main
cluster concentration.

Throughout this paper we will adopt H$_o$ = 100 km s$^{-1}$ Mpc$^{-1}$.

\end{abstract}

\section{Introduction}

Merging clusters are among the most energetic and common
phenomena in the Universe, and the most natural way to
explain the formation of rich clusters of galaxies
within the cold dark matter scenario, which implies 
bottom-up hierarchy of structure formation. The expected 
peculiar velocities in merging clusters
are of the order of 10$^2$ - 10$^3$ km s$^{-1}$.
The merging process generates important perturbations 
in the intracluster medium (ICM), which
leaves signatures in a wide range of frequencies, from 
X--ray energies to the radio band
(see these proceedings). Numerical simulations demonstrate that
merging drives efficient gas transfer to the central regions of
clusters, which could trigger starburst or feed nuclear
radio sources (Bekki 1999).
In this paper we will concentrate
on the relation between merging clusters and their radio
emission properties.

The most important classes of merging-related radio sources
are radio {\it halos} and {\it relics}. They are both
characterised by diffuse extended emission on a scale
up to the order of the megaparsec and have steep 
spectrum ($\alpha \ge 1$) over a wide range of radio frequencies.
Radio halos are typically found in the central regions
of cluster galaxies, while the more elongated relics are usually
located in peripheral cluster regions (Feretti \& Giovannini 1996).

In order to study if and how cluster merging affects the
properties of radio emission of galaxies, and to 
further investigate the relation between mergers and the formation
of radio {\it relics} and radio {\it halos} (see also Feretti,
Giovannini et al., Owen et al., these proceedings), it is 
important to carry out an extensive multifrequency study
of merging clusters. With this aim in mind we have been
carrying out a study in the optical and radio bands of the central 
region of the Shapley Concentration, the most remarkable example
of cluster merging in the nearby Universe, located in the
southern sky at approximately $\alpha_{J2000} = 13^h$  
$\delta_{J2000} = -29^{\circ}$, at
an average distance $<z> \sim 0.03 - 0.05$.

In this paper we will present the first results 
emerging from a radio-optical analysis of the two main
chains of rich clusters in the Shapley Concentration core,
in particular, the A3558 complex and the A3528 complex, hereinafter
A3528-C and A3558-C.
We will briefly describe the observations and the data
reduction, then we will show the radio source counts for both 
cluster complexes 
for comparison with the background source counts and 
we will comment on the results obtained.
Finally we will discuss the astrophysical properties of the  
extended radio source J1324$-$3138, located in A3556 in the
A3558 complex (Venturi et
al. 1998) in the light of the model proposed by Ensslin et. al. (1998)
for the formation of relic sources in merging clusters.

\section{The A3558 and A3528 cluster complexes}

These two cluster complexes can be considered the dynamical
centre of the Shapley Supercluster.

The {\bf A3558} complex (Figure 1) is formed by the three ACO clusters
A3556, A3558 and A3562 and by the two smaller groups SC 1327$-$312 and
SC 1329$-$313 located between the cores of A3558 and A3562.
714 redshifts were obtained by the use of 
multifiber spectroscopic data in this region of the sky (Bardelli
et al. 1994), confirming that A3558-C
is a single physically connected structure, at a mean distance
$<z>$ = 0.0483, almost perpendicular to the
line of sight. The physical connection and merging stage of all clusters in
the chain is emphasised by the X--ray data (Bardelli et al. 1996, Ettori et
al. 1997, Kull \& Boehringer 1998), which shows that the 
distribution of the hot gas in this region remarkably 
follows the distribution  of the optical galaxies. A detailed 
substructure analysis carried out by Bardelli et al. (1998)
showed that the velocity distribution of the galaxies in the
A3558 chain is very complex, with the existence of 
a large number of small groups, further evidence of its
dynamical activity.
We believe that A3558-C is an early stage cluster merger,
after the first core-core encounter.

The {\bf A3528} complex (Figure 2) is formed by the three ACO
clusters A3528, A3530 and A3532 and it is located ad an average
distance $<z>$ = 0.0535. X--ray data (Schindler 1996) show that A3528,
the dominant cluster in this complex, is actually formed by two
distinct groups, called for convenience A3528N and A3558S.
The temperature distribution of the gas suggests that these
two groups are in a pre-merging stage (Schindler 1996). 
The distance between
the centres of A3530 and A3532 is much smaller than the 
Abell radius ($\sim 35^{\prime}$ at this distance), and this
is also evidence of strong interaction.
A dynamical study of A3528-C, based on $\sim$ 600 redshifts in
this region is currently in progress (Bardelli et al. in preparation).
  
\begin{figure}
 \resizebox{\hsize}{!}{\includegraphics{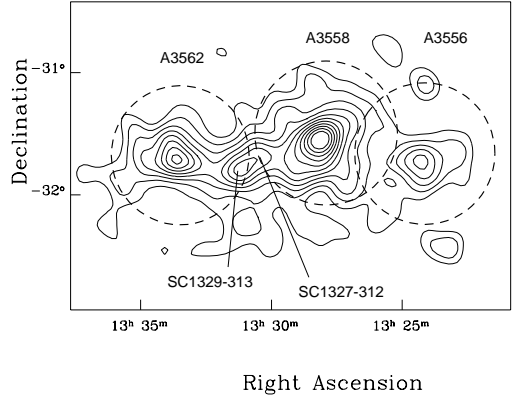}}
\caption[]{Optical isodensity contours of the A3558 complex,
including all galaxies brighter than b$J \le$ 19.5.The dotted
circles represent the Abell radius of the clusters.}
\end{figure}
  
\begin{figure}
 \resizebox{\hsize}{!}{\includegraphics{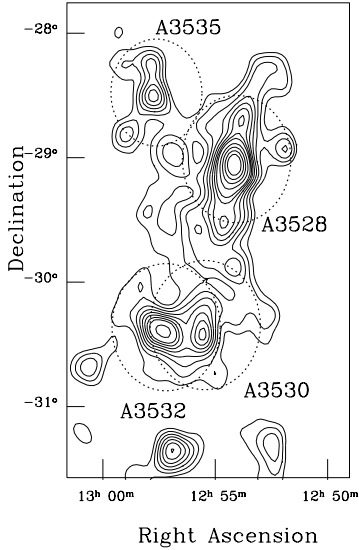}}
\caption[]{Optical isodensity contours of the A3528 complex,
including all galaxies brighter than b$_J \le 19.5$. The
dotted circles represent the Abell radius of the clusters. Filled 
circles represent the pointing centres of the radio observations.}
\end{figure}

\section{The 22 cm radio survey}

\subsection{Observations and data reduction}

In order to pursue the aim of our project and study the
global effects of merging on the radio emission properties
of cluster galaxies and the relation between mergers
and extended {\it relic-} and {\it halo}-type radio sources,
we surveyed A3528-C and A3558-C with
the Australia Telescope Compact Array (ATCA) simultanueously at 
13 cm and 22 cm, with the two array configurations 1.5B and 6C
for a total of 8 hours on each field.
We will concentrate here on the 22 cm results. The resolution
of our observations is $\sim 10^{\prime\prime} \times 5^{\prime\prime}$.
We carried out our observations with a 128 MHz bandwidth, and
in order to minimise the effects of bandwidth smearing in our
large fields of view we 
fully exploited the spectral line mode correlation of the
ATCA using 32 channels. The data were reduced with the MIRIAD
package (Sault, Teuben \& Wright 1995) and the image analysis
was carried out in AIPS. For further details on the observations
and data reduction see Venturi et al. 1997, 1999a and 1999b.
In order to completely cover the sky region occupied by both
complexes we also analysed archive data at the same wavelength
and with comparable resolution (array 1.5C, 2 hours on each field).

The area surveyed in our observations is 1.04 deg$^2$ and 3.25 deg$^2$
for A3528-C and A3558-C respectively.

The noise in our final images ranges from 70$\mu$Jy/beam to 0.2 mJy/beam
therefore we placed a conservative detection limit of 5$\sigma$ for
those fields with highest noise and considered reliable
detections all sources with S$_{22 cm} \ge$ 1 mJy.

\subsection{Results and optical identifications}

The results of our survey are reported in Table 1, where we
give the total number of detected radio sources (Col. 2), the number
and fraction of radio sources with optical counterpart (Col. 3)
and the number and fraction of Shapley radio galaxies (Col. 4).

\begin{table}
      \caption{Results of the 22 cm Survey}
         \label{KapSou}
      \[
          \begin{array}{cccc}
            \hline
            \noalign{\smallskip}
            Complex & N (radio) & N (ID) & N (Shapley) \\
            \noalign{\smallskip}
            \hline
            \noalign{\smallskip}
            A3528 & 152 & 40 (26\%) & 12 (8\%)  \\
            A3558 & 263 & 69 (26\%) & 28 (11\%) \cr
            \noalign{\smallskip}
             \hline
         \end{array}
      \]
   \end{table} 

Among the 12 Shapley radio galaxies in A3528-C, six
exhibit extended morphology, which is a fraction of 50\%. 
In A3558-C only 4/28 of the Shapley radio galaxies are 
extended, i.e. $\sim 14$\% (see Venturi et al. 1999b for further
details on the A3558 complex).
Figure 3 and 4 show the distribution of optical galaxies (dots)
in  A3528-C and A3558-C respectively, with the location
of the radio galaxies superimposed (crosses).

\begin{figure}
 \resizebox{\hsize}{!}{\includegraphics{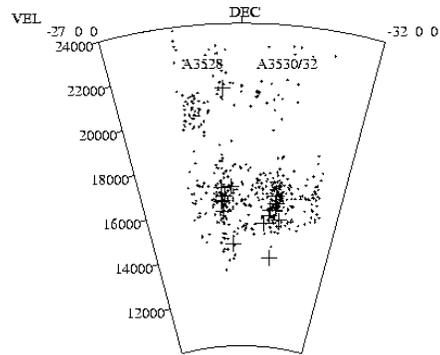}}
\caption[]{Distribution of the A3528 complex in the velocity 
space. The finger of God effect for A3528 and for the
two clusters A3530 and A3532 is evident. Dots represent
the optical galaxies, crosses are the Shapley radio galaxies.}
\end{figure}

\begin{figure}
 \resizebox{\hsize}{!}{\includegraphics{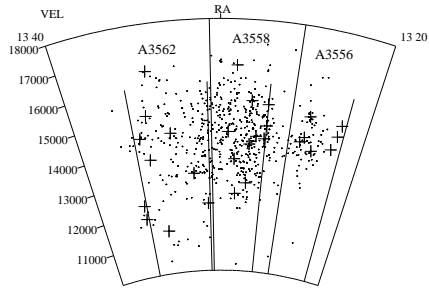}}
\caption[]{Distribution of the A3558 complex in the velocity 
space. The finger of God effect is much less evident here. Dots represent
the optical galaxies, crosses are the Shapley radio galaxies.}

\end{figure}

The six extended
radio galaxies in  A3528-C are all located in the central
regions of the clusters (Reid, Hunstead \& Pierre 1998), 
in particular five of them are in the centre
of A3528N and A3528S (J1254$-$2900, J1254$-$2901, J1254$-$2904,
J1254$-$2913 and J1254$-$2916)
and the sixth (J1257$-$3021) is coincident with the dominant
dumb-bell galaxy in A3532 (see also Gregorini et al. 1994).
A3528-C seems more active than A3558-C at radio wavelengths, and
this could be related to its earlier merging stage. 
The relation between the properties of the extended radio galaxies
in A3528-C and the merging stage will be studied and discussed
in a future work (Venturi et al. in preparation).

The distribution of the extended galaxies in A3558-C 
is remarkably different. Two of them, i.e. J1324$-$3138 and J1333$-$3144,
are located in the central region of A3556 and A3562 respectively
(Venturi et al. 1999a), while the
remaining two, J1322$-$3146 and J1335$-$3153, are located 
respectively at the extreme 
western and eastern end of the whole chain (see Venturi et al. 1997
and Venturi et al. 1999a).

\subsection{Radio Source Counts}

In order to test if the galaxy overdensity and the cluster merger 
in the Shapley Concentration
core reflect in an overdensity in the number of radio
sources, we computed the radio source counts in both complexes
and compared our results to the background source counts
(Prandoni 1997).

Our statistical analysis was carried out using a complete subsample of
all radio sources detected in A3528-C and A3558-C. 
This was necessary to account for sensitivity losses due to the primary
beam attenuation. For the present investigation we selected 
all sources with S $\ge$ 2 mJy within a radius of 17.5 arcmin
from the centre of the respective field. At such distance
the primary beam attenuation of the ATCA at 22 cm is reduced by
a factor of two, therefore sources with a flux density S $\ge$ 2 mJy
are seen as sources with S $\ge$ 1 mJy before the correction.
At the distance of the Shapley Concentration this reflects into
a lower limit on the radio power logP$_{22 cm}$ = 21.7.
The final sample of radio sources used for our statistical
investigation includes 
55 and 144 source for A3528-C and A3558-C.

Our results are shown in Figures 5 and 6 respectively for 
A3528-C and for A3558-C. The line drawn in each
plot represents the radio source counts for the background.
The errors in each flux bin are poissonian.

\begin{figure}
 \resizebox{\hsize}{!}{\includegraphics{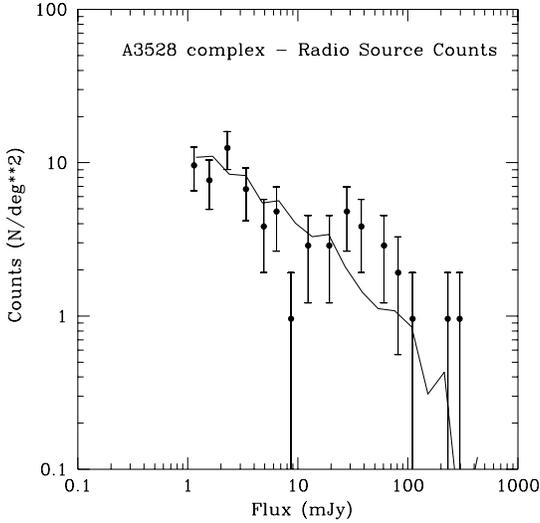}}
\caption[]{Radio source counts for the A3528 complex. The line
drawn shows the distribution of the background source counts (Prandoni 1997).}
\end{figure}

\begin{figure}
 \resizebox{\hsize}{!}{\includegraphics{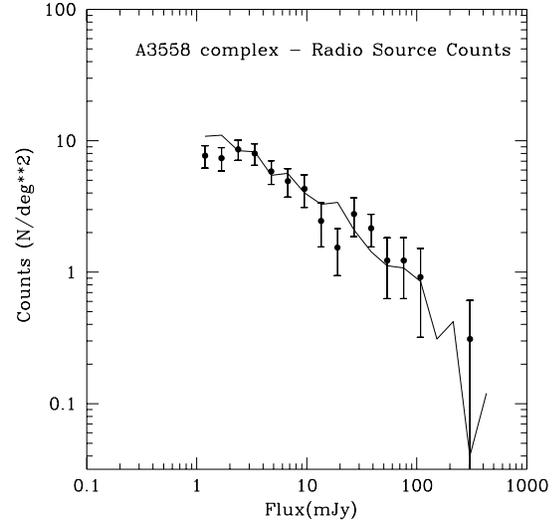}}
\caption[]{Radio source counts for the A3558 complex. The line
drawn shows the distribution of the background source counts (Prandoni 1997).}
\end{figure}

As it is clear from Figures 5 and 6, the distribution of
the source counts in the Shapley Concentration core remarkably
follows that of the background. We remind the reader that
our sample is complete only down to 2 mJy, and the source counts
in the first two bins, lower than the background in both complexes,
reflect our incompleteness in the range 1 - 2 mJy.
A KS test shows that for both complexes the source count
distribution is the same as the background with a 99\% confidence
level.

Our result indicates that even in a remarkably overdense region
as the Shapley Concentration core, the radio source counts
are dominated by background sources. This implies that
the extreme cluster merger in the Shapley Concentration has
no effect on the source counts. Furthermore we can conclude that
cluster merging does not seem to appreciably increase the probability
of a single galaxy to become a radio source over a wide range of
radio power.

\section{J1324$-$3138: a radio relic in A3556?}

It has been recently proposed (Ensslin et al. 1998, Ensslin these 
proceedings) that 
{\it relic} radio sources, trace shock fronts induced by the 
formation of large scale structure through cluster mergers and group 
accretion. Such shock fronts would give rise to electron reacceleration
through magnetic field compression, therefore ``old'' emission
from radio galaxies, faded away because of particle energy
losses, would be ``revived'' during the passage though the shock.
Under this hypothesis the properties of relics and the cluster
gas temperature and pressure would depend on
the accretion shock parameters, such as
the shock radius $r_s$ and the velocity of the infalling  
matter $V_s$.
Their model is able to account for the properties of most
relics known thus far.

\medskip
The extended radio galaxy J1324$-$3138 is located in A3556 (see Figures 1
and 7), one of the ACO clusters in A3558-C and
it was studied in detail
in Venturi et al. (1997). The radio galaxy is located at a projected
distance of $\sim 2.5$ arcmin from the centre of A3556.
Its most important properties can be summarised as follows:

\noindent
{\it (a)} low brigthness tailed radio morphology without jets
and with indication of a very faint nucleus.
The source is associated with a 15.6 mag cluster galaxy belonging 
to a secondary
group in A3556, with $\sigma \sim 222$ km s$^{-1}$ with
respect from the main component in A3556. Its projected 
size is $\sim 182 \times 15$ kpc.

\noindent
{\it (b)} steep spectrum, with $\alpha_{0.8 GHz}^{4.9 GHz} \sim 1.3$;

\noindent
{\it (c)} no polarisation, with a limit on the fractional
polarisation $m < 5$\% up to 4.9 GHz;

\noindent
{\it (d)} low equipartition magnetic field $B_{eq}$ and non-thermal
pressure $P_{nt}$, in particular $B_{eq} = 1.6~~ \mu$G and 
$P_{nt} = 1.5 \times 10^{-13}$ dyn cm$^{-2}$;

\noindent
{\it (e)} age $t\sim 10^8$ yrs.

All these properties led us to the conclusion that J1324$-$3138
is a remnant of a tailed radio galaxy. 

\begin{figure}
 \resizebox{\hsize}{!}{\includegraphics{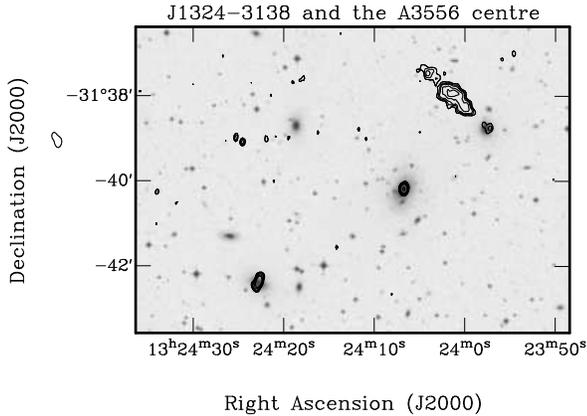}}
\caption[]{Color optical image of the centre of A3556 with
superimposed radio emission at 22 cm. J1324$-$3138 is the
extended source on the top right. The centre of A3556 is
coincident with the cD galaxy at the centre of the image.}
\end{figure}

Given the similarity between this source and the few relics
known up to date, we compared the observed properties
of this source and of the gas in A3556 to the predictions
made by Ensslin et al. (1998) in order to consider the
possibility that J1324$-$3138 is 
a relic source located
on the accretion shock front between the centre of A3556 and the
secondary group visible in the galaxy velocity distribution
for A3556 (Bardelli et al. 1998).
Using the formulas given in Ensslin et al. we obtained for
J1324$-$3138 a shock radius $r_s \sim 1$ Mpc, a viewing angle
$\delta \sim 5^{\circ}$ (defined as the angle between the line
of sight and the line connecting the source and the cluster
centre), and a compression ratio R = 2.9. In Table 2 we give
the results of the comparison between the source properties
and the model.

\begin{table}
      \caption{Comparison with the model}
         \label{KapSou}
      \[
          \begin{tabular}{p{0.5\linewidth}lcc}
            \hline
            \noalign{\smallskip}
            Parameter & Predicted & Observed \\
            \noalign{\smallskip}
            \hline
            \noalign{\smallskip}
            $V_s$ (km s$^{-1}$)    & $\sim 900$ & 936      \\
            P$_2$/P$_1$            & 9.3        & $\sim 5$ \\
            Polarisation           & no~~pol.   & $<$ 5\% \\
            $\sigma$ (km s$^{-1}$) & 217        & 222      \\
            \noalign{\smallskip}
             \hline
         \end{tabular}
      \]
   \end{table} 

\noindent
In Table 2  P$_2$/P$_1$ is the pressure ratio
inside and outside the shock front and $\sigma$ is the velocity 
dispersion of the group, related to the temperature ratio
inside and outside the shock front. The observed values for 
$V_s$ and $\sigma$ are derived from the optical data in Bardelli 
et al. (1998).

The agreement between the predicted and observed critical parameters
in Ensslin model is very good. This analysis 
suggests that the source is located within a small angle
from the cluster centre, and the group it belongs to is
accreting onto A3556. The viewing angle $\delta$ has a 180$^{\circ}$
uncertainty, and the model does not allow us to predict
if the source is located in front or behind the main cluster
condensation. Also, if the source is aligned with the shock front,
the derived geometry suggests that it is almost in the plane of
the sky, so projection effects in its total size are negligible.
This could be a problem, since the dimensions reported in the
literature for the other known relics are considerably larger
(see Feretti \& Giovannini 1996), however with the increasing number
of relic sources being found (these proceedings) the distribution
of linear sizes for this type of sources may be reconsidered in the
future.

The existence of an optical counterpart indicates that the stage of 
nuclear activity, responsible for the production of the 
relativistic electrons and magnetic field 
now ``revived'' by the shock, is fairly recent.
J1324$-$3138 could represent the first stage of a {\it relic} radio source.

A similar multifrequency (radio, spectroscopic, X--ray) analysis
for all extended
sources in the Shapley Concentration core is currently in progress.

\section{Summary and conclusions}

We have surveyed the two chains of merging clusters centered
on A3528 and A2558 located in the central region of the
Shapley Concentration at 22 cm with the Australia Telescope
Compact Array in order to investigate the effects of
merging on their radio emission properties of clusters.
Furthermore we have applied the model proposed by Ensslin et al. (1998)
for the formation of relics to the extended source J1324$-$3138,
located in A3556, in the A3558 complex.

We can summarise our results as follows:

\noindent
{\it (a)} We detected 152 radio sources in A3528-C and
263 radio sources in A3558-C above the flux
detection limit S$\ge$ 1 mJy. The number of
Shapley radio galaxies is respectively 12 (8\%) and 28 (11\%).
In A3528-C we found 6/12 extended radio galaxies, corresponding to
a remarkable
fraction of 50\%. In A3558-C we found 4/28 extended radio galaxies,
i.e. $\sim$ 14\%.

\noindent
{\it (b)} The high optical galaxy density in the cluster merging
environment of the Shapley Concentration does not affect the
radio source counts, which are consistent with the background
source counts. This suggests that merging does not influence
the probability of galaxies to become radio sources.

\noindent
{\it (c)} The properties of J1324$-$3138 are consistent with the
idea that it is a relic source located on the shock front between
the centre of A3556 and the infalling secondary group where its
optical counterpart is located. The viewing angle between the
line of sight and the normal to the shock front is $\delta \sim 5^{\circ}$
and the source lies almost in the plane of the sky.

\begin{acknowledgements}
We wish to thank the organisers for the very fruitful and
successful workshop.
\end{acknowledgements}

\end{document}